\documentclass[aps,pra,twocolumn,showpacs,superscriptaddress,nofootinbib]{revtex4-1}
\usepackage{amsmath}
\usepackage{amssymb}
\usepackage{graphicx}
\usepackage[bookmarks=false]{hyperref}
\usepackage{bm}
\usepackage{times,txfonts}
\usepackage{ntheorem}
\usepackage{enumerate}
\usepackage{enumitem}
\usepackage{mathrsfs}
\usepackage{epstopdf}
\usepackage{subfigure}
\hypersetup{colorlinks=true,citecolor=blue,linkcolor=red,urlcolor=blue,pdfstartview=FitH,bookmarksopen=true}

\DeclareMathOperator{\Tr}{Tr}

\newcommand{\bra}[1]{\langle #1\rvert}
\newcommand{\ket}[1]{\lvert #1\rangle}
\newcommand{\abs}[1]{\lvert #1\rvert}

\def\CC{{\rm\kern.24em \vrule width.04em height1.46ex depth-.07ex \kern-.30em C}}
\def\RR{{\rm\kern.24em \vrule width.04em height1.46ex depth-.07ex
\kern-.30em R}}
\def\P{{\rm I\kern-.25em P}}

\begin{document}
\title{Universal freezing of asymmetry}
\author{Da-Jian Zhang}
\affiliation{Department of Physics, Shandong University, Jinan 250100, China}
\affiliation{School of Mathematics, Shandong University, Jinan 250100, China}
\author{Xiao-Dong Yu}
\affiliation{Department of Physics, Shandong University, Jinan 250100, China}
\author{Hua-Lin Huang}
\affiliation{School of Mathematical Sciences, Huaqiao University, Quanzhou 362021, China}
\author{D. M. Tong}
\email{tdm@sdu.edu.cn}
\affiliation{Department of Physics, Shandong University, Jinan 250100, China}
\date{\today}
\begin{abstract}
Asymmetry of quantum states is a useful resource in applications such as quantum metrology, quantum communication, and reference frame alignment. However, asymmetry of a state tends to be degraded in physical scenarios where environment-induced noise is described by covariant operations, e.g., open systems constrained by superselection rules, and such degradations weaken the abilities of the state to implement quantum information processing tasks. In this paper, we investigate under which dynamical conditions asymmetry of a state is totally unaffected by the noise described by covariant operations. We find that all asymmetry measures are frozen for a state under a covariant operation if and only if the relative entropy of asymmetry is frozen for the state. Our finding reveals the existence of  universal freezing of asymmetry, and provides a necessary and sufficient condition under which asymmetry is totally unaffected by the noise.
\end{abstract}
\maketitle

Symmetry is a central concept in quantum mechanics, describing invariant features of a quantum system with respect to the action of a group of transformations \cite{Messiah1962}. For a specific symmetry, two relevant notions are asymmetric states and covariant operations, which are the states that break the symmetry and the quantum operations that respect the symmetry, respectively. In the physical world, all elementary interactions are expected to have specific symmetries \cite{Marvian2008}. For example, the interactions that do not have preferred direction are rotationally invariant and hence have SO(3) symmetry. The presence of a symmetry in a system generally imposes restrictions on the manipulation of the system, which results in nontrivial limitations on the implementation of quantum information processing tasks.  Interestingly, asymmetric states can be exploited to overcome the restrictions and allow one to implement quantum information processing tasks that would otherwise be forbidden \cite{Bartlett2007}. For example, in the presence of a conservation law, it is forbidden to measure exactly an observable that does not commute with a conserved quantity, but it is still possible to measure approximatively the observable with the aid of asymmetric states \cite{Wigner1952,Araki1960}. Asymmetry of states is a useful resource for implementing quantum information processing tasks \cite{Bartlett2007}, and
the exploitation of asymmetric states has been carried out in applications, such as quantum metrology \cite{Lloyd2004,Lloyd2006,Giovannetti2011,Hall2012}, quantum communication \cite{Duan2001,Gisin2007}, and reference frame alignment \cite{Peres2001,Bagan2001,Chiribella2004}.

By taking asymmetry as a physical
resource, a resource theory of asymmetry, just like the resource theory of entanglement \cite{Horodecki2009}, has been recently developed. The abilities of an asymmetric state to overcome the restriction imposed by a symmetry are analogous to the abilities of an entangled state to overcome the restriction of local operations and classical communication (LOCC). Asymmetric states and covariant operations in the asymmetry theory correspond respectively to entangled states and LOCC in the entanglement theory, or resource states and free operations in a general resource theory \cite{Brandao2015}. Over the recent years, a lot of effort has been devoted to the formulation of a unified and quantitative theory of asymmetry, which aims to quantify the abilities of asymmetric states for implementing quantum information processing tasks \cite{Vaccaro2008,Gour2008,Gour2009,Toloui2011,Skotiniotis2012,Marvian2013,Marvian2014,MarvianPRA2014,MarvianPRA20141,Marvian2016,Yadin2016,Piani2016,Marvian20161}.
The asymmetry theory was first used to measure the quality of a quantum reference frame \cite{Vaccaro2008,Gour2008,Gour2009}, subsequently exploited to find the consequences of symmetries for open quantum dynamics \cite{Marvian2013,Marvian2014,MarvianPRA2014,MarvianPRA20141}, and recently linked to the resource theory of coherence \cite{Marvian2016,Yadin2016,Piani2016,Marvian20161}.
It turns out that the asymmetry theory is applicable to a wide spectrum of physical contexts, and based on them, a number of asymmetry measures, such as the unique asymptotic measure of frameness \cite{Gour2008}, the relative entropy of frameness \cite{Gour2009}, the Holevo asymmetry measure \cite{Marvian2014}, the Wigner-Yanase skew information \cite{Marvian2016}, and the quantum Fisher information \cite{Yadin2016}, have been proposed.

Asymmetry of a state is a useful resource for quantum information processing, but it may suffer from degradation arising from the interaction between the system and its environment. Indeed, in many physical scenarios, e.g., all open systems constrained by superselection rules, the environment-induced noise is described by covariant operations \cite{Bartlett2003}, whose actions tend to destroy asymmetry of the state and hence make the state less useful for implementing quantum information processing tasks.  A challenge in exploiting the resource is therefore to preserve asymmetry of states from the degradation caused by the covariant noise, i.e., the noise described by covariant operations. When the asymmetry of a state is frozen, i.e., remains constant, the ability of the state to implement some quantum information processing task will not be weakened if the ability exploited in the task is based on the frozen asymmetry measure. However, there are many different asymmetry measures, each of which is used to quantify one ability of a state to implement a different quantum information processing task, and one asymmetry measure being frozen does not imply other asymmetry measures being frozen, too. The covariant noise may not weaken some abilities of a state if these abilities are based on the frozen asymmetry measures, but it can still weaken the other abilities that are based on unfrozen asymmetry measures. Only the state with universal freezing of asymmetry can keep all the abilities of asymmetry resource totally unaffected by the covariant noise. Here, by the phrase, universal freezing of asymmetry, we mean that the asymmetry of a state is frozen regardless of asymmetry measures adopted, i.e., all asymmetry measures of the state are frozen under a certain covariant operation. The question then is as follows: under which dynamical conditions does the universal freezing of asymmetry occur for a state under a covariant operation? This is an important issue, since only in this case asymmetry of a state is totally unaffected by the covariant noise.  In this paper, we address this issue. We will show that all asymmetry measures are frozen for a state under a covariant operation if and only if the relative entropy of asymmetry is frozen for the state.

Note that similar issues on how to preserve other resources of a state from the degradation caused by noise have been widely addressed.  For instance, the preservation of entanglement, the freezing of quantum correlations, and the freezing of quantum coherence were investigated in Refs. \cite{An2010,Yi2010,Franco2015,Franco2016,Yi2016}, Refs. \cite{Mazzola2010,Mazzola2011,You2012,An2013,Aaronson2013,Chanda2015,Cianciaruso2015}, and Refs. \cite{Bromley2015,Yu2016,Silva2016}, respectively.  The present investigation aims to fill the gap in the resource theory of asymmetry.

To present our finding clearly, it is instructive to specify some notions, such as symmetric states, asymmetric states, covariant operations, and asymmetry measures.

Consider a quantum system equipped with a Hilbert space $\mathcal{H}$. Let $G$ be a group of physical transformations acting on $\mathcal{H}$ through a unitary representation $U_g$. The group $G$ together with its unitary representation specifies the symmetry under consideration. We represent the transformation associated with the group element $g$ by the map $\mathcal{U}_g$, i.e., $\mathcal{U}_g(\rho)=U_g\rho U_g^\dagger$.

A state $\delta$ is said to be a symmetric state with respect to $G$ if
\begin{eqnarray}
\mathcal{U}_g(\delta)=\delta,\nonumber
\end{eqnarray}
for all $g\in G$. The set of all symmetric states is denoted by $\mathcal{S}$. All other states are called asymmetric states with respect to $G$. Hereafter, we use $\rho$ to represent a general state, and $\delta$ specially to denote a symmetric state.

A quantum operation $\Lambda$ is said to be a covariant operation with respect to $G$ if
\begin{eqnarray}\label{cov.ope.}
\Lambda\circ\mathcal{U}_g=\mathcal{U}_g\circ\Lambda,
\end{eqnarray}
for all $g\in G$. That is, the transformation realized by applying first $\mathcal{U}_g$ and then $\Lambda$ is equivalent to that realized by applying first $\Lambda$ and then $\mathcal{U}_g$.

A functional $A$ mapping states to real numbers can be taken as an asymmetry measure if it satisfies the following two conditions:\\
(i) $A(\rho)\geq 0$ for all states $\rho$, and $A(\rho)=0$ if $\rho\in\mathcal{S}$;\\
(ii) $A(\rho)\geq A(\Lambda(\rho))$ for all covariant operations $\Lambda$, that is, $A$ is non-increasing under covariant operations.

One of the asymmetry measures is the relative entropy of asymmetry $A_r$ \cite{Note1}. It is defined as
\begin{eqnarray}\label{Ar}
A_r(\rho)=\min_{\delta\in\mathcal{S}}S(\rho\|\delta),
\end{eqnarray}
where $S(\rho\|\delta)=\Tr\rho(\log\rho-\log\delta)$ is the quantum relative entropy. In the case that $G$ is a finite or compact Lie group, this measure admits a closed-form expression \cite{Gour2009},
\begin{eqnarray}\label{clo.form}
A_r(\rho)=S(\rho\|\Lambda_{G}(\rho))=S(\Lambda_{G}(\rho))-S(\rho),
\end{eqnarray}
where $S(\rho)=-\Tr(\rho\log\rho)$ is the von Neumann entropy, and $\Lambda_{G}$ is the $G$-twirling operation, defined as $\Lambda_{G}(\rho)=\int_{G}dgU_g\rho U_g^\dagger$ with the integral being performed over the Haar measure. Note that for a finite group, there is $\Lambda_{G}(\rho)=\frac{1}{\abs{G}}\sum_{g\in G}U_g\rho U_g^\dagger$ with $\abs{G}$ being the order of the group.

With these notions, we can now state our main finding as a theorem.

\textit{Theorem.} $A(\rho_t)=A(\rho_0)$ for all asymmetry measures $A$ if and only if $A_r(\rho_t)=A_r(\rho_0)$, where $\rho_t=\Lambda_t(\rho_0)$ with $\Lambda_t$ being a covariant operation and $\rho_0$ being an initial state.

We only need to prove that $A(\rho_t)=A(\rho_0)$ if $A_r(\rho_t)=A_r(\rho_0)$, since $A_r$ is certainly frozen if all asymmetry measures are frozen.

First, we show that $S(\Lambda_t(\rho_0)\|\Lambda_t(\delta_0))=S(\rho_0\|\delta_0)$, where $\delta_0$ denotes the symmetric state achieving the minimum in the expression $A_r(\rho_0)=\min_{\delta\in\mathcal{S}}S(\rho_0\|\delta)$. By definition,
\begin{eqnarray}\label{pf1}
A_r(\rho_0)=S(\rho_0\|\delta_0).
\end{eqnarray}
Since the quantum relative entropy is contracting under completely positive and trace-preserving (CPTP) maps \cite{Lindblad,Uhlmann}, we have
\begin{eqnarray}\label{pf2}
S(\Lambda_t(\rho_0)\|\Lambda_t(\delta_0))\leq S(\rho_0\|\delta_0).
\end{eqnarray}
On the other hand, as $\Lambda_t$ is a covariant operation mapping symmetric states to symmetric states, we have $\Lambda_t(\delta_0)\in\mathcal{S}$ and hence
\begin{eqnarray}\label{pf3}
A_r(\rho_t)=\min_{\delta\in\mathcal{S}}S(\Lambda_t(\rho_0)\|\delta)\leq S(\Lambda_t(\rho_0)\|\Lambda_t(\delta_0)).
\end{eqnarray}
Combining Eqs. (\ref{pf1}), (\ref{pf2}), and (\ref{pf3}), we obtain
\begin{eqnarray}\label{pf4}
A_r(\rho_t)\leq S(\Lambda_t(\rho_0)\|\Lambda_t(\delta_0))\leq S(\rho_0\|\delta_0)=A_r(\rho_0).
\end{eqnarray}
In the condition of $A_r(\rho_t)=A_r(\rho_0)$, Eq. (\ref{pf4}) leads to
\begin{eqnarray}\label{pf5}
A_r(\rho_t)=S(\Lambda_t(\rho_0)\|\Lambda_t(\delta_0)),
\end{eqnarray}
and
\begin{eqnarray}\label{pf6}
S(\Lambda_t(\rho_0)\|\Lambda_t(\delta_0))=S(\rho_0\|\delta_0).
\end{eqnarray}
Equation (\ref{pf5}) implies that $\Lambda_t(\delta_0)$ is the symmetric state achieving the minimum in the expression $A_r(\rho_t)=\min_{\delta\in\mathcal{S}}S(\rho_t\|\delta)$, while Eq. (\ref{pf6}) shows that the equality for the contractivity of quantum relative entropy in Eq. (\ref{pf2}) is attained.

In passing, we would like to point out that if the symmetry group $G$ is restricted to finite or compact Lie groups, the above proof for Eq. (\ref{pf6}) can be simplified by resorting to Eq. (\ref{clo.form}). Indeed, from Eq. (\ref{clo.form}), it follows that $A_r(\rho_0)=S(\rho_0\|\Lambda_G(\rho_0))$ and $A_r(\rho_t)=S(\rho_t\|\Lambda_G(\rho_t))$. Noting that $\delta_0=\Lambda_G(\rho_0)$ and $\Lambda_G\circ\Lambda_t=\Lambda_t\circ\Lambda_G$, we have $\Lambda_G(\rho_t)=\Lambda_G\circ\Lambda_t(\rho_0)=\Lambda_t\circ\Lambda_G(\rho_0)=\Lambda_t(\delta_0)$.
Hence, there is $A_r(\rho_0)=S(\rho_0\|\delta_0)$ and $A_r(\rho_t)=S(\Lambda_t(\rho_0)\|\Lambda_t(\delta_0))$. Equation (\ref{pf6}) then follows immediately from the condition  $A_r(\rho_t)=A_r(\rho_0)$.


Second, we demonstrate that there exists a covariant operation $R_t$ such that $R_t(\rho_t)=\rho_0$ and $R_t(\delta_t)=\delta_0$, where $\delta_t=\Lambda_t(\delta_0)$. Hereafter, we use $\delta_t$ to represent $\Lambda_t(\delta_0)$ for simplicity. Let
\begin{eqnarray}\label{Kraus-rep}
\Lambda_t(\rho)=\sum_nK_n(t)\rho K_n^\dagger(t)
\end{eqnarray}
be the Kraus representation of $\Lambda_t$, where $K_n(t)$ are the Kraus operators satisfying $\sum_n K_n^\dagger(t) K_n(t)=I$. From the celebrated result about the contractivity of quantum relative entropy \cite{Petz,Hayden}, it follows that Eq. (\ref{pf6}) is valid if and only if there exists a CPTP map $R_t$ such that
\begin{eqnarray}\label{pf7}
R_t(\rho_t)=\rho_0~~\textrm{and}~~ R_t(\delta_t)=\delta_0.
\end{eqnarray}
In the case that $\delta_t$ is invertible, $R_t$ can be given explicitly by the formula \cite{Hayden}
\begin{eqnarray}\label{pf-R}
R_t(\rho)=\sum_n \delta_0^{\frac{1}{2}}K_n^\dagger(t)\delta_t^{-\frac{1}{2}}\rho\delta_t^{-\frac{1}{2}} K_n(t)\delta_0^{\frac{1}{2}}.
\end{eqnarray}
We therefore only need to prove that the CPTP map expressed by Eq. (\ref{pf-R}) is covariant.
For convenience, we rewrite Eq. (\ref{pf-R}) as follows,
\begin{eqnarray}\label{pf8}
R_t=R_1\circ R_2\circ R_3,
\end{eqnarray}
where $R_1(\rho)=\delta_0^{\frac{1}{2}}\rho\delta_0^{\frac{1}{2}}$, $R_2(\rho)=\sum_nK_n^\dagger(t)\rho K_n(t)$, and
$R_3(\rho)=\delta_t^{-\frac{1}{2}}\rho\delta_t^{-\frac{1}{2}}$.
In order to prove that Eq. (\ref{pf-R}) defines a covariant operation, it suffices to show that each $R_i$, $i=1,2,3$, fulfills Eq. (\ref{cov.ope.}), i.e., $R_i\circ\mathcal{U}_g=\mathcal{U}_g\circ R_i$, for all $g\in G$. Since $\delta_0$ is a symmetric state, there is $\mathcal{U}_g(\delta_0)=\delta_0$, i.e., $[\delta_0,U_g]=0$. This implies that $\delta_0$ and $U_g$ are simultaneously diagonalizable, which further implies that $\delta_0^{\frac{1}{2}}$ and $U_g$ are simultaneously diagonalizable, because $\delta_0^{\frac{1}{2}}$ shares common eigenvectors with $\delta_0$. Hence, $[\delta_0^{\frac{1}{2}}, U_g]=0$. It follows that $R_1\circ\mathcal{U}_g(\rho)=\delta_0^{\frac{1}{2}}U_g\rho U_g^\dagger\delta_0^{\frac{1}{2}}=U_g\delta_0^{\frac{1}{2}}\rho \delta_0^{\frac{1}{2}}U_g^\dagger=\mathcal{U}_g\circ R_1(\rho)$. That is, $R_1$ fulfills Eq. (\ref{cov.ope.}). Similarly, we can prove that $R_3$ fulfills Eq. (\ref{cov.ope.}). Now, it remains to show that $R_2$ fulfills Eq. (\ref{cov.ope.}), too. By definition, there is $\mathcal{U}_g\circ\Lambda_t=\Lambda_t\circ\mathcal{U}_g$. As an immediate consequence, the equality  $\Tr[X\mathcal{U}_g\circ\Lambda_t(Y)]=\Tr[X\Lambda_t\circ\mathcal{U}_g(Y)]$ holds for any operators $X$ and $Y$. Inserting the explicit expressions of $\mathcal{U}_g$ and $\Lambda_t$ into this equality, we have $\Tr\{XU_g[\sum_nK_n(t)YK_n^\dagger(t)]U_g^\dagger\}=\Tr\{X[\sum_nK_n(t)U_gYU_g^\dagger K_n^\dagger(t)]\}$. Since the trace of a matrix is cyclic, i.e., $\Tr(AB)=\Tr(BA)$, for two arbitrary matrices $A$ and $B$, we have $\Tr\{[\sum_nK_n^\dagger(t) U_g^\dagger XU_gK_n(t)]Y\}=\Tr\{U_g^\dagger[\sum_n K_n^\dagger(t) XK_n(t)]U_gY\}$. Noting that $U_g^\dagger=U_{g^{-1}}$ and $U_g=U_{g^{-1}}^\dagger$, we further have $\Tr[R_2\circ\mathcal{U}_{g^{-1}}(X)Y]=\Tr[\mathcal{U}_{g^{-1}}\circ R_2(X)Y]$. Since this equality holds for any operators $X$ and $Y$, there is $R_2\circ\mathcal{U}_{g^{-1}}=\mathcal{U}_{g^{-1}}\circ R_2$.
Letting $g$ in this equation run over all elements of $G$, we then have  $R_2\circ\mathcal{U}_{g}=\mathcal{U}_{g}\circ R_2$, for all $g\in G$, which means that $R_2$ fulfills Eq. (\ref{cov.ope.}). Therefore, Eq. (\ref{pf-R}) defines a covariant operation satisfying Eq. (\ref{pf7}). In the case that $\delta_t$ is not invertible, instead of Eq. (\ref{pf-R}), $R_t$ should be expressed as
\begin{eqnarray}\label{pf10}
R_t(\rho)=\sum_n\delta_0^{\frac{1}{2}}K_n^\dagger(t)\delta_t^{-\frac{1}{2}}\rho\delta_t^{-\frac{1}{2}} K_n(t)\delta_0^{\frac{1}{2}}+P\rho P,
\end{eqnarray}
where $P$ is the orthogonal projector onto the kernel of $\delta_t$, and $\delta_t^{-\frac{1}{2}}$ is defined to be the square root of the Moore-Penrose pseudoinverse of $\delta_t$, i.e., $\delta_t^{-\frac{1}{2}}=\sum_{\lambda_i\neq0}\lambda_i^{-\frac{1}{2}}\ket{\phi_i}\bra{\phi_i}$, provided that the spectral decomposition of $\delta_t$ reads $\delta_t=\sum_i\lambda_i\ket{\phi_i}\bra{\phi_i}$. Similarly, one can show that Eq. (\ref{pf10}) defines a covariant operation satisfying Eq. (\ref{pf7}).

Third, with the foregoing arguments, it is ready to show the conclusion $A(\rho_t)=A(\rho_0)$. By combining the two covariant operations $\Lambda_t$ and $R_t$, there is
\begin{equation}
  \rho_0\xrightarrow{\Lambda_t}\rho_t\xrightarrow{R_t}\rho_0.
  \label{eq:rho}
\end{equation}
Since all the asymmetry measures $A$ are non-increasing under covariant operations, Eq. (\ref{eq:rho}) results in
\begin{equation}
  A(\rho_0)\ge A(\rho_t)\ge A(\rho_0),
  \label{eq:rhoineq}
\end{equation}
which implies that $A(\rho_t)=A(\rho_0)$. This completes the proof of the theorem.

Our theorem provides a necessary and sufficient condition under which asymmetry of a state is totally unaffected by the covariant noise. It is applicable to all kinds of symmetry groups. When the symmetry group is a finite or compact Lie group, the relative entropy of asymmetry admits a closed-form expression in Eq. (\ref{clo.form}), and consequently our theorem provides a computable criterion for identifying the states with universal freezing of asymmetry. All the states with universal freezing of asymmetry can be obtained by solving the equation $A_r(\rho_t)=A_r(\rho_0)$, although it may be difficult to solve analytically this equation to obtain all the solutions since the calculation of entropy is complicated. However, it is generally unnecessary to obtain all the solutions. In practical applications, researchers are
usually interested only in some special states. In this case, one only needs to examine the desired states, to which our theorem is quite useful.

In the following, we demonstrate the usefulness of our theorem by presenting two examples, of which one is about time evolution of an open system and another is about measurement on a system.

\textit{Example 1: time evolution.}--Consider the time evolution of an open system subject to a superselection rule \cite{Bartlett2003}. For simplicity, we suppose that the system is composed of two qubits and the superselection rule is associated with the group $U(1)$ with the unitary representation $U_\theta=\exp[i\theta(\sigma_z\otimes I+I\otimes\sigma_z)]$, where $\sigma_z=\ket{0}\bra{0}-\ket{1}\bra{1}$. The model can be generalized to the multiqubit case. As discussed in Ref. \cite{Bartlett2003}, the superselection rule restricts the allowed dynamics of the system to those that are covariant with respect to the associated group. Given this situation, we consider the time evolution of the system described by the following covariant operation,
\begin{eqnarray}\label{ex2-map}
\Lambda_t(\rho)=(1-p)\rho+p\sigma_z\otimes\sigma_+\rho\sigma_z\otimes\sigma_-+p\sigma_z\otimes\sigma_-\rho\sigma_z\otimes\sigma_+,\nonumber\\
\end{eqnarray}
where $\sigma_+=\ket{1}\bra{0}$, $\sigma_-=\ket{0}\bra{1}$, and $0\leq p\leq 1$ is a parameter dependent on the time $t$.

Let us examine a family of pure states, expressed as
\begin{eqnarray}\label{ex2-state}
\ket{\varphi_0}=\lambda_0\ket{00}+\lambda_1\ket{10},
\end{eqnarray}
where $\lambda_m$ are complex numbers satisfying $\abs{\lambda_0}^2+\abs{\lambda_1}^2=1$. By using our theorem, we now show that all asymmetry measures are frozen forever for any initial state $\ket{\varphi_0}$ under the covariant operation $\Lambda_t$. To this end, we only need to show that $A_r(\rho_t)$ is a constant, where $\rho_t=\Lambda_t(\rho_0)$ with $\Lambda_t$ being defined in Eq. (\ref{ex2-map}) and $\rho_0=\ket{\varphi_0}\bra{\varphi_0}$ being an initial state.

Direct calculations show that
\begin{eqnarray}
\rho_t&=&(1-p)(\lambda_0\ket{00}+\lambda_1\ket{10})(\lambda_0^*\bra{00}+\lambda_1^*\bra{10})\nonumber\\
&&+p(\lambda_0\ket{01}-\lambda_1\ket{11})(\lambda_0^*\bra{01}-\lambda_1^*\bra{11}).\nonumber
\end{eqnarray}
Noting that the $G$-twirling operation is $\Lambda_G(\rho)=\sum_{i=0}^2P_i\rho P_i$, where $P_0=\ket{00}\bra{00}$, $P_1=\ket{01}\bra{01}+\ket{10}\bra{10}$, and $P_2=\ket{11}\bra{11}$, we further have
\begin{eqnarray}
\Lambda_G(\rho_t)&=&(1-p)\left(\abs{\lambda_0}^2\ket{00}\bra{00}+\abs{\lambda_1}^2\ket{10}\bra{10}\right)\nonumber\\
&&+p\left(\abs{\lambda_0}^2\ket{01}\bra{01}+\abs{\lambda_1}^2\ket{11}\bra{11}\right).\nonumber
\end{eqnarray}
We can then obtain the relative entropy of asymmetry,
\begin{eqnarray}\label{ex2-fre}
A_r(\rho_t)&=&S(\rho_t\|\Lambda_G(\rho_t))=S(\Lambda_G(\rho_t))-S(\rho_t)\nonumber\\
&=&\sum_{m=0}^1\big[-(1-p)\abs{\lambda_m}^2\log\left((1-p)\abs{\lambda_m}^2\right)-p\abs{\lambda_m}^2\nonumber\\
&&\times\log\left(p\abs{\lambda_m}^2\right)\big]+(1-p)\log(1-p)+p\log p\nonumber\\
&=&\sum_{m=0}^1-\abs{\lambda_m}^2\log \abs{\lambda_m}^2=A_r(\rho_0).
\end{eqnarray}
Equation (\ref{ex2-fre}) shows that $A_r(\rho_t)$ is a constant. This implies that all asymmetry measures manifest freezing forever for the two-qubit system initially in the state expressed by Eq. (\ref{ex2-state}) undergoing the time evolution defined by Eq. (\ref{ex2-map}). That is, universal freezing of asymmetry occurs in this case.

\textit{Example 2: measurement process.}-- Consider the measurement process discussed first in Ref. \cite{Bartlett20061}. Although we have thus far focused on the preservation of asymmetry from the degradation caused by environment-induced noise, our theorem is also applicable to the case where the degradation of asymmetry arises from a measurement process as long as the process can be described by covariant operations.

As shown in Refs. \cite{Bartlett20061,Poulin2007,Boileau2008}, performing a covariant measurement has some back reaction on a quantum reference frame and degrades asymmetry of the reference frame. In order to demonstrate the degradation of a phase reference frame, the authors of Ref. \cite{Bartlett20061} considered the estimation task of measuring the phase of a large number of qubits relative to a single phase reference frame. Their estimation task consists of a sequence of covariant measurements, which are performed on the combined systems consisting of the reference frame and each qubit, one after another. As a result of these repeated measurements, the reference frame becomes more and more useless for implementing their estimation task. It implies that the asymmetry of the reference frame is degraded \cite{Bartlett20061}. In the following, we show that if the reference frame is initially prepared in some special states, the asymmetry of the reference frame manifests universal freezing for a certain number of measurements and is therefore totally unaffected by these measurements.

The authors of Ref. \cite{Bartlett20061} used an oscillator mode to act as the phase reference frame. The Hilbert space of the phase reference frame is then the Fock space spanned by $\{\ket{n}$, $n=0,1,2,\dots\}$. The symmetry under consideration is the group $U(1)$ with the unitary representation $U_\theta=\exp(i\theta N)$, where $N$ is the number operator. Accordingly, the $G$-twirling operation is $\Lambda_G(\rho)=\sum_{n=0}^\infty\ket{n}\bra{n}\rho\ket{n}\bra{n}$, whose effect is to project onto eigenvectors of the group generator $N$. As a result of a single measurement, the state of the phase reference frame is updated by the covariant operation \cite{Bartlett20061},
\begin{eqnarray}
\Lambda(\rho)=\frac{1}{2}\rho+\frac{1}{4}\ket{0}\bra{0}\rho\ket{0}\bra{0}+\frac{1}{4}A^\dagger\rho A+\frac{1}{4}A\rho A^\dagger,
\end{eqnarray}
where $A=\sum_{n=0}^\infty\ket{n}\bra{n+1}$. In this situation, the time index  $t$ is simply an integer specifying the number of measurements that have taken place. The state of the phase reference frame following the $t$-th measurement is $\rho_t=\Lambda(\rho_{t-1})$, with $\rho_0$ denoting the initial state prior to any measurement.

We examine a family of pure states, expressed as
\begin{eqnarray}\label{ex-state}
\lambda_0\ket{N}+\lambda_1\ket{3N}+\cdots+\lambda_M\ket{(2M+1)N},
\end{eqnarray}
where $\lambda_m$ are complex numbers satisfying $\sum_{m=0}^M\abs{\lambda_m}^2=1$, and $N$ and $M$ are positive integers. Hereafter, we use $\ket{\varphi_n}$ to denote the state $\lambda_0\ket{N+n}+\lambda_1\ket{3N+n}+\cdots+\lambda_M\ket{(2M+1)N+n}$ for simplicity. Then, the state in Eq. (\ref{ex-state}) can be simply written as $\ket{\varphi_0}$. By using our theorem, we show that if the number of measurements is less than $N$, the asymmetry of the phase reference frame initially in the state expressed by Eq. (\ref{ex-state}) manifests universal freezing. To this end, we only need to show that the relative entropy of asymmetry $A_r(\rho_t)$ are constants, where $\rho_0=\ket{\varphi_0}\bra{\varphi_0}$ and $t<N$.

By detail calculations, we obtain
\begin{eqnarray}\nonumber
\rho_t=\sum_{n=-t}^{t} p_n(t)\ket{\varphi_n}\bra{\varphi_n},
\end{eqnarray}
with
\begin{eqnarray}\nonumber
p_n(t)=\sum_{n\leq k\leq\frac{t+n}{2}}\binom{t}{k}\binom{t-k}{k-n}\left(\frac{1}{2}\right)^{t-n+2k},
\end{eqnarray}
where $k$ represents an nonnegative integer and $\binom{\cdot}{\cdot}$ denotes the binomial coefficient, i.e., $\binom{t}{k}=\frac{t!}{k!(t-k)!}$. Further calculations show that
\begin{eqnarray}\nonumber
\Lambda_G(\rho_t)=\sum_{n=-t}^t\sum_{m=0}^Mp_n(t)\abs{\lambda_m}^2\ket{(2m+1)N+n}\bra{(2m+1)N+n}.
\end{eqnarray}

With the aid of the above expressions, we can obtain the
relative entropy of asymmetry,
\begin{align}
  &A_r(\rho_t)=S(\rho_t\|\Lambda_G(\rho_t))=S(\Lambda_G(\rho_t))-S(\rho_t)\notag\\
  =&-\sum_{n=-t}^t\sum_{m=0}^Mp_n(t)\abs{\lambda_m}^2\log\left[p_n(t)\abs{\lambda_m}^2\right]+\sum_{n=-t}^tp_n(t)\log p_n(t)\notag \\
  =&-\sum_{m=0}^M\abs{\lambda_m}^2\log\abs{\lambda_m}^2=A_r(\rho_0).\label{ex1-fre}
\end{align}
Equation (\ref{ex1-fre}) shows that the relative entropy of asymmetry for
the state $\rho_t$ is constant. Therefore, all asymmetry measures
manifest freezing for the phase reference frame initially in the
state expressed by Eq. (\ref{ex-state}).
Universal freezing of asymmetry occurs in this case, too.

In conclusion, we have investigated the freezing phenomenon of asymmetry and put forward a theorem on this issue. It shows that all measures of asymmetry are frozen for a state under a covariant operation if and only if the relative entropy of asymmetry is frozen for the state. This theorem is applicable to all kinds of covariant operations defined by all groups, including but not limited to finite and compact Lie groups.
Our finding reveals the existence of universal freezing of asymmetry, and more importantly provides a necessary and sufficient condition under which asymmetry of a state is totally unaffected by the covariant noise. Note that similar issues about other resources such as quantum correlations and quantum coherence have been widely addressed, and the freezing phenomenon of correlations and the freezing phenomenon of coherence have already been found. Our investigation fills a gap in the resource theory of asymmetry.

This work was supported by the China Postdoctoral Science Foundation (Grant No. 2016M592173). X.D.Y. acknowledges support from the National Natural Science Foundation of China (Grant No. 11575101). H.L.H. acknowledges support from the National Natural Science Foundation of China (Grant No. 11571199). D.M.T. acknowledges support from the National Basic Research Program of China (Grant No. 2015CB921004).

\end{document}